\documentclass[rmp,twoside]{revtex4}

\usepackage{graphicx}
\usepackage{epstopdf}
\usepackage{algorithm}
\usepackage{algpseudocode}
\usepackage{subcaption}
\usepackage{amsmath}

\newcommand{\ilast}[0]{m}
\newcommand*\LabelSymbol{\ensuremath{\boldsymbol\ell}}
\newcommand*\EmptyLabel{\ensuremath{\boldsymbol\ell_\mathbf{\emptyset}}}
\newcommand*\emptyLabel{\ensuremath{\ell_\emptyset}}

\begin{document}

\title{An Automorphic Distance Metric and its Application to Node Embedding for Role Mining}

\newcommand{\citicaffiliation}{Research Center for Information and Communications Technologies, University of Granada, Spain}

\newcommand{\urgaffiliation}{Department of Computer Science and Artificial Intelligence, University of Granada, Spain}

\author{V\'ictor Mart\'inez}
\email{victormg@acm.org}
\affiliation{\citicaffiliation}

\author{Fernando Berzal}
\email{berzal@acm.org}
\affiliation{\urgaffiliation}

\author{Juan-Carlos Cubero}
\email{jc.cubero@decsai.ugr.es}
\affiliation{\urgaffiliation}

\begin{abstract}
Role is a fundamental concept in the analysis of the behavior and function of interacting entities represented by network data. Role discovery is the task of uncovering hidden roles. Node roles are commonly defined in terms of equivalence classes, where two nodes have the same role if they fall within the same equivalence class. Automorphic equivalence, where two nodes are equivalent when they can swap their labels to form an isomorphic graph, captures this common notion of role. The binary concept of equivalence is too restrictive and nodes in real-world networks rarely belong to the same equivalence class. Instead, a relaxed definition in terms of similarity or distance is commonly used to compute the degree to which two nodes are equivalent. In this paper, we propose a novel distance metric called automorphic distance, which measures how far two nodes are of being automorphically equivalent. We also study its application to node embedding, showing how our metric can be used to generate vector representations of nodes preserving their roles for data visualization and machine learning. Our experiments confirm that the proposed metric outperforms the RoleSim automorphic equivalence-based metric in the generation of node embeddings for different networks.
\end{abstract}


\maketitle

\section{Introduction}
Role discovery is defined as the process of finding sets of nodes following similar connectivity patterns or structural behaviors \citep{rossi2015role}. The role of a node can be understood as the function that node plays in the network. Different studies have shown the importance of roles in different domains, including predator-prey food webs  \citep{luczkovich2003defining}, international relations \citep{hafner2009network}, or the function of proteins in proteomes \citep{holme2005role}.

Unfortunately, this problem has received limited attention when compared to community detection \citep{lancichinetti2009community, fortunato2010community,papadopoulos2012community}, despite the fact that role discovery identifies complementary information and has found application in several useful network data mining tasks. For example, roles can be used to model and characterize the behaviors of entities in a network to predict structural changes and detect anomalies \citep{rossi2013modeling}. Since the same roles can be observed across different networks, this information has been successfully exploited for transfer learning \citep{henderson2012rolx}. Role information can also be used for enhanced visualization of interesting patterns in graphs \citep{xing2010state}. Additional applications of role discovery are covered in more detail in \citep{rossi2015role}.

Formally, two nodes have the same role if, given an equivalence relation, they belong to the same equivalence class \citep{white1976social, burt1990detecting}. Different equivalence classes haven been studied for nodes in networks.

Structural equivalence, where two nodes play the same role if they are connected to exactly the same neighbor nodes, has been widely studied \citep{lorrain1971structural, sailer1978structural}. These nodes will have exactly the same topological properties, such as degree, clustering coefficient, or centrality, since they are indistinguishable from a structural point of view. However, different authors have pointed out the limitations of structural equivalence for modeling roles or positions (the name that roles receive in sociology), since structural equivalence is more related to the concept of locality than the actual concept of role \citep{borgatti1992notions}. If the constraint of needing to be connected to exactly the same neighbors is relaxed to being connected to neighbors with exactly the same topological function, we obtain automorphic equivalence classes, where two nodes are equivalent if they can swap their labels to form an isomorphic graph \citep{pattison1988network, friedkin1997social}. Automorphically equivalent nodes will also have exactly the same topological properties but, without the requirement of locality imposed by structural equivalence, pairs of nodes at distances larger than two can still have the same role. Therefore, automorphic equivalence is more closely related to the intuitive concept of role, understood as the function of a node within a network.

Other equivalence classes, less relevant than the ones previously mentioned, are not covered in this work. Regular equivalence deserves special mention due to its importance as a relaxation of automorphic equivalence that only requires being connected to nodes with the same function, omitting the actual count of connections \citep{everett1994regular}. Regular equivalence does not preserve topological properties and is more suited to hierarchically-organized networks \citep{luczkovich2003defining}.

These binary equivalences are strict mathematical abstractions that rarely occur in real-world networks, leading to all nodes being assigned a different role. In practice, these equivalences are relaxed to similarities, allowing two nodes to play the same role by partially satisfying the constraints imposed by the mathematical definition of structural, automorphic, or regular equivalence.

In this paper, we present a novel automorphic distance metric, capturing distances between nodes in terms of automorphic equivalence. According to the network structure, two nodes will be at a distance that is proportional to how far they are from being automorphically equivalent. This leads to a softer definition of automorphic roles, instead of forcing all nodes to fit in strict classes of roles. However, when needed, these distances can be used to discover and instantiate specific role classes. Our distance function satisfies metric axioms, as we prove below, does not require external parameters nor feature engineering, and is computable for nodes across different networks. We also present different applications of our proposal, with special emphasis on generating node embeddings that preserve node roles. Node embeddings are vector representations of nodes capturing relevant information in terms of pair-wise distances \citep{goyal2017graph}. Much work has been done in embedding techniques that preserve neighborhoods or communities \citep{grover2016node2vec,zheng2016node}, but, to the best of our knowledge, our work is the first one on role-preserving embeddings.

Our paper is structured as follows. In Section 2, we discuss the relevant related work. In Section 3, we describe our proposed automorphic distance metric and study its admissibility as a distance metric, as well as its computational complexity. In Section 4, we analyze its application to the generation of node embeddings that preserve roles and show how it outperforms previously proposed approaches. Finally, conclusions and suggestions for future research are presented in Section 5.

\section{Related Work}
Different metrics have been proposed to measure node similarity. One of the most popular metrics is SimRank \citep{jeh2002simrank}, which iteratively computes similarity scores based on the hypothesis that two nodes are similar if they link to similar nodes. Different extensions of SimRank have been proposed \citep{hamedani2016simrank}. SimRank recursively computes similarity of two nodes according to the average similarity of all their neighbor pairs, which can also be interpreted, as suggested by its original authors, as how soon two random walkers will meet if they start from these nodes. Thus, this definition is not suitable as a metric of similarity capturing automorphic equivalence because it requires the two nodes to be close to play the same role. Other similarity measures not based on SimRank have been proposed, such as PageSim \citep{lin2006pagesim} and Leicht's vertex similarity \citep{leicht2006vertex}. These similarities are formally rejected as valid metrics for capturing automorphic equivalence in \citep{jin2014scalable}.

Since automorphic equivalence ensures the same topological properties, some authors have tried to capture automorphic equivalence by defining a similarity function over a set of network topological properties \citep{chen2015fast}. The problem of these feature-based methods is that they require combining different complex hand-crafted features obtained by experts, which is far from trivial in practice. In addition, they cannot guarantee which set of features will correctly approximate automorphic equivalence, leading to a very limited approach to automorphic equivalence discovery.

\begin{figure}[!ht]
\centering
\includegraphics[width=0.35\textwidth]{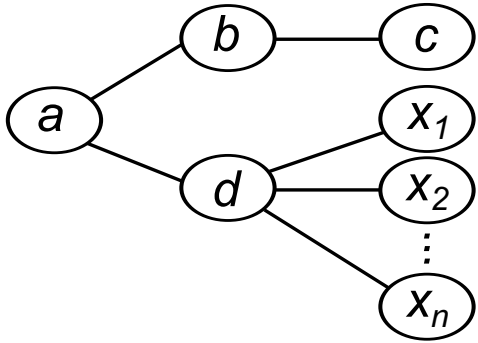} 
\caption{Example graph where RoleSim yields inconsistent values. Node $d$ has a one to many relation to $x_i$ nodes.}
\label{fig:rolesim_fail}
\end{figure}

As far as we know, RoleSim \citep{jin2011axiomatic,jin2014scalable} is the only proposed metric that tries to formally capture the concept of automorphic equivalence without using limited approximations based on hand-crafted topological features. Omitting the decay factor they introduce, by setting it to $0$ in order to capture the global network topology, this similarity measure is iteratively computed until convergence as

$$s(x,y) = \max_{M(x,y)} \frac{\sum_{(u,v) \in M(x,y)} s(u,v)}{deg(x)+deg(v)-\vert M(x,y) \vert}$$

\noindent where $deg(n)$ is the degree of a node $n$ and $M(x,y)$ is the optimal assignment of nodes in the neighborhood of $x$ to nodes in the neighborhood of $y$ maximizing the expression; that is, the pairs of neighbors of $x$ and $y$ with maximal similarity. In their work, Jin et al. prove that this function satisfies the axioms required to be considered a valid role similarity metric. RoleSim is a form of generalized Jaccard coefficient based on a recursive definition of the similarity of neighbor roles. Despite the admissibility of RoleSim, their approach presents several limitations. The RoleSim similarity can be considered an automorphic distance by taking its complementary or Jaccard distance: $d(x,y)=1-s(x,y)$. The problem is that the Jaccard coefficient is a normalized metric, which leads to a normalized distance. As will be shown in our experimentation, this normalization has a negative impact on the results obtained by RoleSim. In addition, this similarity function exhibits serious inconsistencies. For example, in the graph shown in Figure \ref{fig:rolesim_fail}, where node $d$ has a one-to-many relationship to $x_i$ nodes, the node pair $(a,c)$ has the same exact similarity than any pair $(a,x_i)$, independently of the number of $x_i$ nodes. This simple example shows the limitations of RoleSim when trying to capture automorphic similarity.

As far as we know, no distance metric has been proposed that is able to capture the concept of automorphic distance in a consistent way, without relying on approximations based on extracted topological properties nor forcing the normalization of the distance function.

\section{A Novel Automorphic Distance Metric}
An isomorphism is a bijection between the nodes of two graphs where two nodes are adjacent in one graph if and only if the nodes that result from applying the bijective function are also adjacent in the other graph. An automorphism is an isomorphism from one graph to itself. Therefore, two nodes are automorphically equivalent if there exists an automorphism creating a correspondence between them.

One form of testing for automorphic equivalence is computing the canonical form of graphs. Graph canonization is the task of computing a labeling for nodes in a graph such that every isomorphic graph yields the same canonical labeling. Given a canonized graph, two automorphically-equivalent nodes must have been assigned the same label. As previously stated, automorphic equivalence is too restrictive to appear in real-world networks, leading to most nodes having different canonical labels.

The solution that we propose to this problem is the definition of distances between labels, which ultimately allows the definition of distances between nodes based on the concept of automorphic equivalence. This distance will be proportional to the number of changes that need to be done in the network to transform one label or equivalence class into another. A zero distance implies that two nodes are automorphically equivalent and play exactly the same role. According to this distance $d$, we can say that nodes $x$ and $y$ are more automorphically similar or have a more similar role than $u$ and $v$ if $d(x,y)<d(u,v)$. In order to propose a valid distance metric, we must also prove that our metric satisfies the distance metric axioms.

\begin{figure*}[!ht]
\centering
\begin{subfigure}[t]{0.49\textwidth}
\centering
\includegraphics[width=0.85\textwidth]{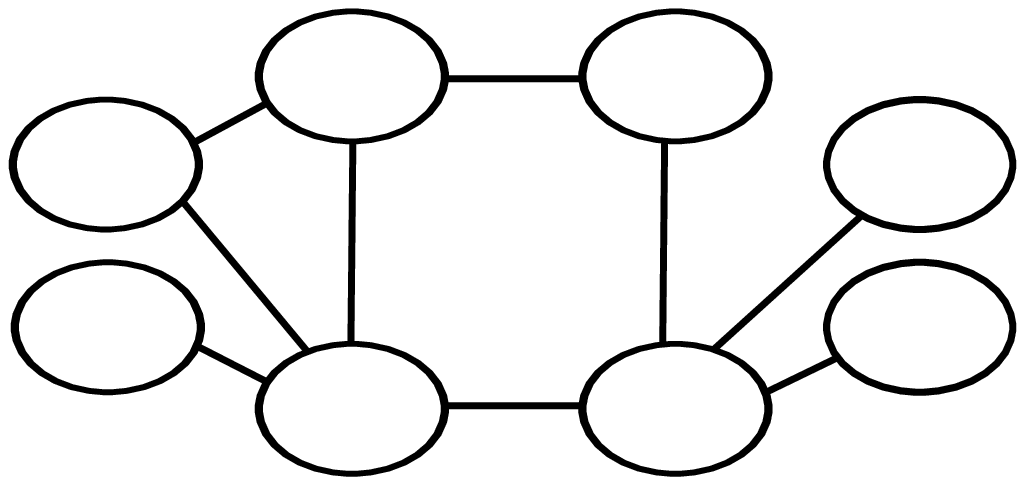}
\caption{Original graph.}
\label{fig:net1}
\end{subfigure}
\begin{subfigure}[t]{0.49\textwidth}
\centering
\includegraphics[width=0.85\textwidth]{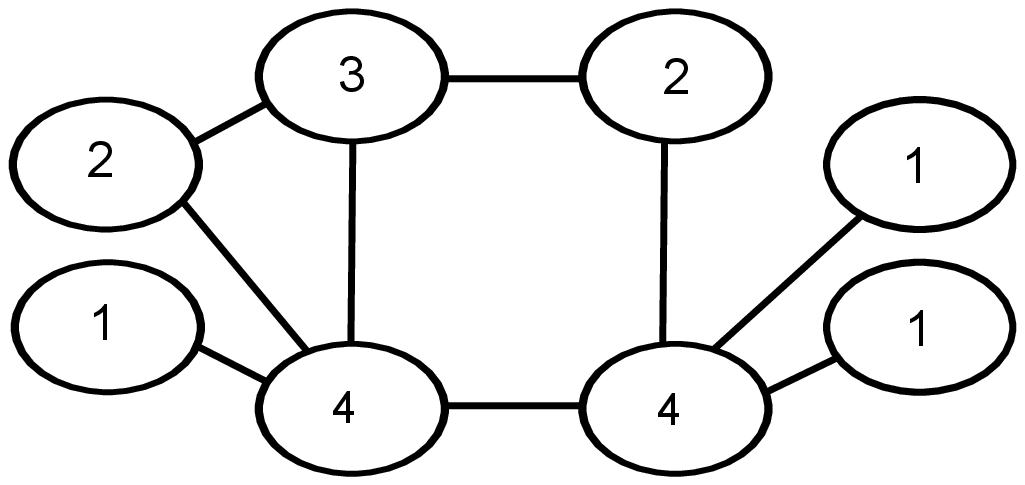}
\caption{Initial labeling.}
\label{fig:net2}
\end{subfigure}
\vspace{0.5cm}

\begin{subfigure}[t]{0.49\textwidth}
\centering
\includegraphics[width=0.85\textwidth]{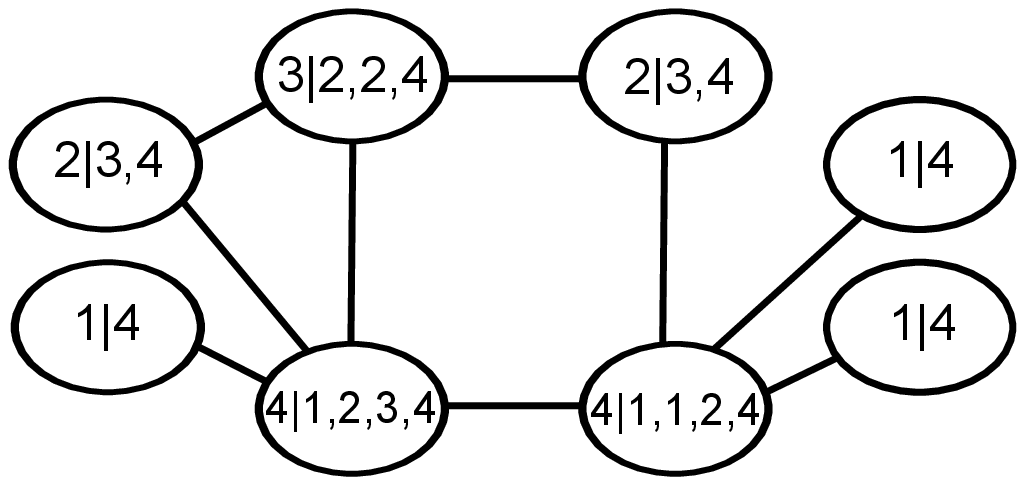}
\caption{First iteration.}
\label{fig:net3}
\end{subfigure}
\begin{subfigure}[t]{0.49\textwidth}
\centering
\includegraphics[width=0.85\textwidth]{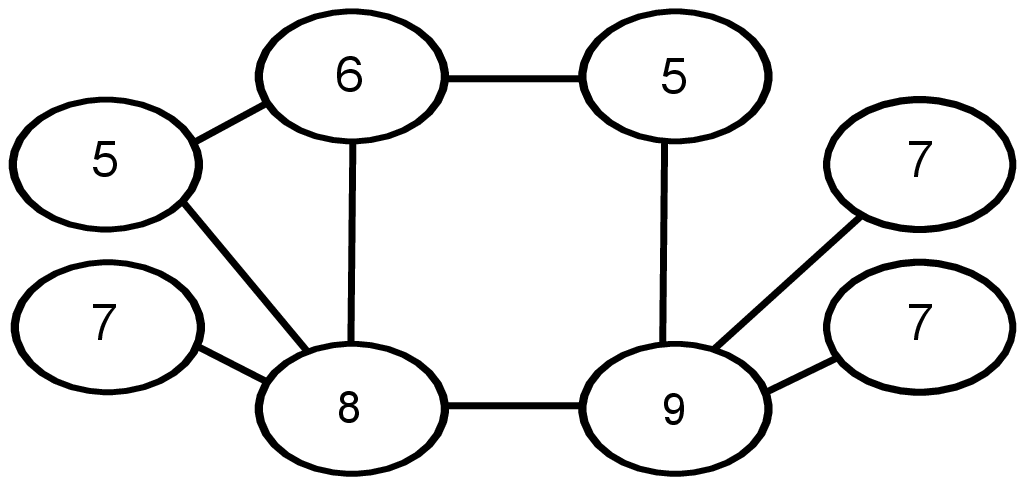}
\caption{First relabeling.}
\label{fig:net4}
\end{subfigure}
\vspace{0.5cm}

\begin{subfigure}[t]{0.49\textwidth}
\centering
\includegraphics[width=0.85\textwidth]{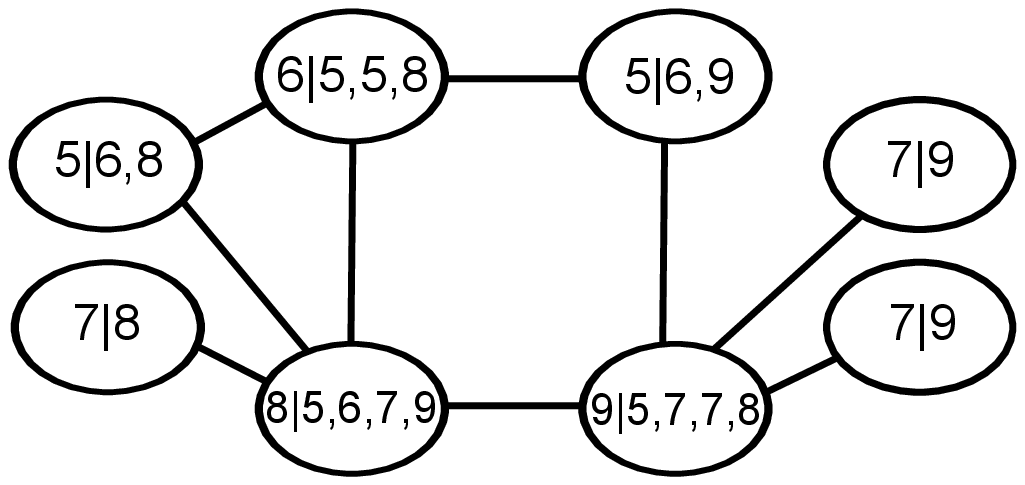}
\caption{Second iteration.}
\label{fig:net5}
\end{subfigure}
\begin{subfigure}[t]{0.49\textwidth}
\centering
\includegraphics[width=0.85\textwidth]{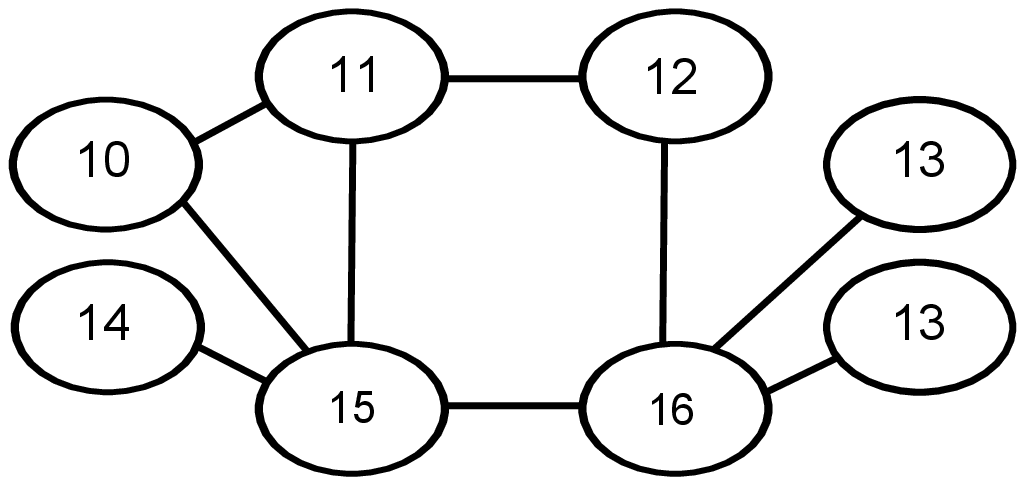}
\caption{Second and final relabeling.}
\label{fig:net6}
\end{subfigure} 
\caption{The Weisfeiler-Lehman canonization algorithm applied to a simple graph.}
\label{fig:weisfeiler_lehman}
\end{figure*}

Our work is based on the 1-dimensional Weisfeiler-Lehman test of isomorphism \citep{weisfeiler1968reduction, furer2001weisfeiler}, also known as color refinement, which is an algorithm to compute the canonical labeling of graphs. These canonical labels can be used to solve related problems, such as the computation of efficient graph kernels \citep{shervashidze2011kernels}. The Weisfeiler-Lehman algorithm works by initially assigning a label to each node according to its degree, so nodes with the same degree have the same initial label. Then, the algorithm iteratively updates these labels by the following procedure. First, it takes the labels from neighbor nodes, concatenates them according to certain arbitrary order (the same ordering must be applied for all nodes), and finally appends the label of the node at the beginning of the obtained list. Each different sequence is substituted by a newly generated unique label so nodes exhibiting exactly the same sequence are assigned the same label. This refinement process is repeated until labels stabilize, that is, when every pair of nodes with the same label in the previous iteration have the same label in the current iteration. Therefore, after $m$ iterations, which depend on the network diameter, the canonical form is achieved and an additional iteration is required for testing the stabilization condition. These final labels are the canonical form of the graph and, therefore, two nodes with the same final label are automorphically equivalent. An example of running the algorithm in a simple graph is shown in Figure \ref{fig:weisfeiler_lehman}.

It can be noted that some pairs of the labels appearing in the same iteration of the Weisfeiler-Lehman algorithm are more similar than others. The automorphic distance between two nodes can be defined as the distance between their canonical labels. We propose a scheme to compute distances between the labels that are obtained by the Weisfeiler-Lehman algorithm. Since distances are only defined for labels appearing in the same iteration, a special label associated to nodes of degree $0$, which we call the empty label $\ell_{\emptyset}$, is considered for convenience. Isolated nodes are directly assigned this label and left out of the iterative process.

Since labels created in the initial assignment are based on node degree, we define the distance of labels of nodes $x$ and $y$ as the number of links that must be added to or removed from node $x$ to transform it into node $y$. This can be easily computed as their absolute degree difference:

\begin{equation}
d(\ell_0(x),\ell_0(y)) = \vert deg(x)-deg(y) \vert,
\label{eq:initdist}
\end{equation}
where $\ell_0(n)$ is the initially-assigned label to node $n$. This definition of distance for initial labels is also valid for isolated nodes, with degree $0$, which have been assigned the empty label.

Given these distances for initial labels, the distance between labels for the subsequent iterations can be computed as the distance of the optimally-matched pairs of labels of their neighbors from the previous iteration. The distance of labels from the $i$-th iteration can be computed as

\begin{equation}
d(\ell_{i}(x),\ell_{i}(y))=
\min_{M_{i-1}(x,y)} \sum_{(u,v) \in M_{i-1}(x,y)} d(\ell_{i-1}(u),\ell_{i-1}(v)),
\label{eq:subdist}  
\end{equation}

\noindent where $M_{i-1}(x,y)$ is the optimal assignment of neighbors of $x$ to neighbors of $y$ that minimizes the expression and, therefore, it is just the sum of distances between neighbors of $x$ and $y$. If the neighborhood of one node is larger than the neighborhood of the other, leading to unmatched nodes, these nodes are directly matched with virtual nodes, which are labeled with $\ell_{\emptyset}$. The distances of unmatched nodes to the empty label can be seen as the cost, in terms of distance, of inserting and transforming a virtual isolated node to obtain a node with the label of the unmatched node. The optimal assignment, which would consider $O(n!)$ alternatives using a naive brute force approach, can be computed in polynomial time using the Hungarian algorithm \citep{kuhn1955hungarian}.

Initially, Equation \ref{eq:initdist} and, in subsequent iterations, Equation \ref{eq:subdist} are used to compute a distance table. At any given time, only distances from two iterations need to be maintained: the distances currently being computed and the distances from the most recent previous iteration.

The described iterative process is carried out for each iteration of the 1-dimensional Weisfeiler-Lehman algorithm until label stabilization. The automorphic distance between a pair of nodes is  defined as the distance between their canonical labels. It should be noted that labels can be represented using any set of symbols. However, for simplicity, we represent labels as positive integers.

\begin{algorithm}
\caption{Automorphic Distance Algorithm}\label{alg:autodist}
\begin{algorithmic}[0]
\Procedure{Automorphic Distance}{}\newline
\textbf{Input:} Set of nodes $N$ of an undirected graph.
\State Initialize $i=0, stabilized=false, r = HashTable()$.
\State Set $\ell_i(x) = deg(x)$ $\forall x\in N$.
\State Set $d(\ell_i(x),\ell_i(y)) = \vert deg(x) - deg(y) \vert$ $\forall x,y \in N \times N$.
\While{\textbf{not} $stabilized$}
\State Set $\ilast=i$. Iterate $i=i+1$.
\State Set $h_i(x) = concatenate(sort(neighbors(x)))$.
\State Set $c_i(x) = concatenate(\ell_{i-1}(x), h_i(x))$.
\State Set $\ell_i(x) = unique(c_i(x))$.
\State Use Hungarian algorithm to compute $M_{i-1}(x,y)$ $\forall x,y \in N \times N$.
\State Set $d(\ell_{i}(x),\ell_{i}(y))=
\min_{M_{i-1}(x,y)} \sum_{(u,v) \in M_{i-1}(x,y)} d(\ell_{i-1}(u),\ell_{i-1}(v))$ $\forall x,y \in N \times N$.

\State Set $stabilized = true$.
\State Store in $r$ the label of $x$ in iteration $i-1$ $\forall x\in N$ as $r[\ell_{i}(x)] = \ell_{i-1}(x)$.
\State If the entry $r[\ell_{i}(x)]$ was already set with a different label, set $stabilized = false$. 
\EndWhile
\State Set $d(x,y) = d(\ell_{\ilast}(x),\ell_{\ilast}(y))$ for each pair of nodes $x,y \in N \times N$.
\newline
\textbf{Output:} Pairwise automorphic distances $d(x, y)$ for each pair of nodes $x,y \in N \times N$.
\EndProcedure
\end{algorithmic}
\end{algorithm}

The complete algorithm is shown in Algorithm \ref{alg:autodist}. The function $neighbors(x)$ returns the set of neighbor nodes of node $x$. The function $sort(s)$ sorts a set of elements. The ordering among elements is not relevant for the algorithm, but the same ordering must always be applied.  The function $concatenate(x_1,\ldots,x_n)$ returns the concatenation of elements $x_1,\ldots,x_n$. Finally, the function $unique(s)$ generates and returns a unique symbol, such an integer, for each observed unique string $s$, where $unique(s)=unique(s')$ if and only if $s=s'$.


\subsection{Case Example}
In this section, we show an illustrative example applying the proposed automorphic distance metric to the network shown in Figure \ref{fig:net1}. 

The proposed algorithm for computing the automorphic distance initially assigns a label to each node according to its degree, as shown in Figure \ref{fig:net2}. Therefore, two nodes will have the same label if and only if they have the same degree. The initial distance table, represented as an upper triangular matrix due to the symmetry of distances, as will be proved in Section \ref{sec:symmetry}, is shown in Table \ref{tab:dist1}. This distance table is computed using Equation \ref{eq:initdist} according to the initial label assignations. For example, the distance between the labels $1$ and $4$ is $3$, since this value is the absolute degree difference of the corresponding nodes.

\begin{table}[!h]
    \centering
        \begin{tabular}{|c|c|c|c|c|c|}
        	\hline
            \LabelSymbol$/$\LabelSymbol & \EmptyLabel & \textbf{1} & \textbf{2} & \textbf{3} & \textbf{4}\\\hline
           \EmptyLabel & 0 & 1 & 2 & 3 & 4\\\hline
           \textbf{1} &  & 0 & 1 & 2 & 3\\\hline
            \textbf{2} & &  & 0 & 1 & 2\\\hline
            \textbf{3} & &  &  & 0 & 1\\\hline
            \textbf{4} & &  &  &  & 0\\\hline
        \end{tabular}
        \caption{Initialization of the distance table.}
       \label{tab:dist1}
       \end{table}

After the initialization, the algorithm enters into its main loop and performs its first iteration. For each node, the labels of its neighbors are ordered and concatenated with its own label, as shown in Figure \ref{fig:net3}. For example, the only node with label $3$ has the associated string $3 \vert 2,2,4$, since its neighbors have labels $2$, $4$, and $2$. These concatenated strings are replaced by a new label, chosen so that two nodes are assigned the same new label if and only if they had the same concatenated string. This process generates a new labeling as shown in Figure \ref{fig:net4}. Given these new labels, the algorithm computes their pairwise distances using Equation \ref{eq:subdist}, which are shown in Table \ref{tab:dist2}.

    \begin{table}[!h]
        \centering
        \begin{tabular}{|c|c|c|c|c|c|c|}
        	\hline
           \LabelSymbol$/$\LabelSymbol & \EmptyLabel & \textbf{5} & \textbf{6} & \textbf{7} & \textbf{8} & \textbf{9}\\\hline
           \EmptyLabel &0&7&8&4&10&8\\\hline
           \textbf{5} &&0&3&3&3&3\\\hline
            \textbf{6}& &  &0 &4 &4 &2\\\hline
            \textbf{7}& &  & &0 & 6&4\\\hline
            \textbf{8} & &  &  & &0 &2\\\hline
            \textbf{9} & &  &  & & &0\\\hline
        \end{tabular}
        \caption{Distance table after the first relabeling.}
        \label{tab:dist2}
    \end{table}
    
For example, in order to compute the distance between labels $5$ and $9$, the optimal assignment between their neighbors minimizing the summation of distances, according to the previous iteration, must be obtained. In the previous iteration, $3$ and $4$ were the labels of the neighbors of nodes with label $5$. Likewise, $1$, $1$, $2$, and $4$ were the labels of the neighbors of nodes with label $9$. The Hungarian algorithm matches these neighbor labels to minimize their sum of distances: $(3,2)$ and $(4,4)$ according to Table \ref{tab:dist1}. Since the two neighbor labels $1$ of label $9$ were left unmatched, they are both matched with the empty label as $(1,\emptyLabel)$. Given this optimal assignment, we can compute the distance between labels $5$ and $9$ as:
$$ d_1(5,9) = d_0(3,2)+d_0(4,4)+d_0(1,\emptyLabel)+d_0(1,\emptyLabel) = 1+0+1+1 = 3 $$

Following this iterative process, the algorithm performs the second iteration. Concatenated strings are computed as shown in Figure \ref{fig:net5} and labels are updated as shown in Figure \ref{fig:net6}. It can be easily seen that labels have stabilized, obtaining the canonical labeling of this graph. The stabilization condition can be tested by performing an additional iteration and observing that nodes with label $13$ are assigned the same label, while the other nodes are assigned an unique new label. This new labeling would be equivalent to the labeling obtained in this iteration, the condition required to achieve stabilization. The pairwise distances computed in this iteration are shown in Table \ref{tab:dist3}.

    \begin{table}[!h]
    \centering
        \begin{tabular}{|c|c|c|c|c|c|c|c|c|}
        	\hline
           \LabelSymbol$/$\LabelSymbol & \EmptyLabel & \textbf{10} & \textbf{11} & \textbf{12} & \textbf{13} & \textbf{14} & \textbf{15} & \textbf{16}\\\hline
           \EmptyLabel &0&18&24&16&8&10&27&25\\\hline
           \textbf{10} &&0&10&2&12&8&13&11\\\hline
           \textbf{11} &&&0&12&16&14&13&7\\\hline
           \textbf{12} &&&&0&8&10&11&13\\\hline
           \textbf{13} &&&&&0&2&19&17\\\hline
           \textbf{14} &&&&&&0&21&15\\\hline
           \textbf{15} &&&&&&&0&6\\\hline
           \textbf{16} &&&&&&&&0\\\hline
        \end{tabular}
        \caption{Distance table after the second and final relabeling.}
        \label{tab:dist3}
        \end{table}

For instance, let us compute the distance between labels $11$ and $16$. We start by finding the optimal assignment that minimizes the pairwise distances of their neighbor labels in the previous iteration, which are $5,5,$ and $8$ for label $11$ and $5,7,7,$ and $8$ for label $16$. The Hungarian algorithm obtains the optimal matching $(5,5), (5,7), $ and $(8,8)$, with an additional $(\emptyLabel,7)$, due to the difference of the node degrees associated to labels $11$ and $16$. Once this optimal matching has been obtained, we can easily compute the distance between labels $11$ and $16$ using Equation \ref{eq:subdist} as:

$$ d_2(11,16) = d_1(5,5)+d_1(5,7)+d_1(8,8)+d_1(\emptyLabel,7) = 0+3+0+4 = 7 $$

The automorphic distance between a pair of nodes is defined as the distance between their canonical labels, which are the final labels assigned in the iteration where stabilization is achieved. Therefore, in our example, the distance between nodes is given by Table \ref{tab:dist3}. For example, we can see how nodes with canonical labels $13$ and $14$ are close to being automorphically equivalent, since their automorphic distance is only $2$. In contrast, nodes with canonical labels $14$ and $15$ have a large automorphic distance, equal to $21$, which indicates that they are far from being automorphically equivalent.

\subsection{Metric Admissibility}
In this section, we prove that the distance function that we have defined is a valid metric or distance function. In order to assert this statement, we must prove the following four conditions: non-negativity, identity of indiscernibles, symmetry, and triangle inequality.

To prove these conditions, we note that Equation \ref{eq:subdist} can be recursively decomposed as

\begin{align}
\begin{split}\label{eq:decomposition}
d(\ell_{\ilast}(x),\ell_{\ilast}(y))&=
\min_{M_{\ilast-1}(x,y)} \sum_{(u,v) \in M_{\ilast-1}(x,y)} d(\ell_{\ilast-1}(u),\ell_{\ilast-1}(v))\\
&=\min_{\substack{M_{\ilast-1}(x,y)\\ M_{\ilast-2}(u,v)}} \sum_{\substack{(u,v)\in\\\ M_{\ilast-1}(x,y)}} \sum_{\substack{(u',v')\in\\ M_{\ilast-2}(u,v)}} d(\ell_{\ilast-2}(u'),\ell_{\ilast-2}(v'))\\
&=\ldots\\
&=\sum_{(x',y') \in M'(x,y)} d(\ell_{0}(x'),\ell_{0}(y'))\\
&=\sum_{(x',y') \in M'(x,y)} \vert deg(x')-deg(y') \vert \\
\end{split}
\end{align}

\noindent where $M'(x,y)$ is the set of pairs of nodes that appear in the recursive summation at the deepest level of recursion as a result of choosing the optimal assignment in each iteration.

\subsubsection{Proof of Non-Negativity}
Non-negativity requires that the distance function satisfies $d(x,y) \geq 0$ for any possible pair of nodes $x$ and $y$. Given the decomposition of our metric as shown in Equation \ref{eq:decomposition}, it is straightforward to see that the summation of absolute values is guaranteed to be always equal or greater than $0$.

\subsubsection{Proof of the Identity of Indiscernibles}
The identity of indiscernibles implies that the distance function satisfies $d(x,y) = 0$ if and only if $x \equiv y$. The Weisfeiler-Lehman algorithm guarantees that automorphically equivalent pairs of nodes are assigned the same canonical label, and non-automorphically equivalent pairs of nodes are assigned different canonical labels.

Equation \ref{eq:decomposition} is only equal to zero when $deg(x')=deg(y')$ for every pair of nodes in $M'(x,y)$. Two nodes can only have the same canonical label if they are automorphically equivalent, as guaranteed by the Weisfeiler-Lehman algorithm. If two nodes are assigned the same canonical label, their neighbors must have been assigned equivalent labels in all the iterations. Since the distance of a label to itself is $0$, we can see that the summation yields $0$, leading to a zero distance for nodes when $x \equiv y$.

On the other side, when two nodes have different canonical labels, their neighbors must have been assigned different labels during the execution of the algorithm. This implies that, in the recursive decomposition shown in Equation \ref{eq:decomposition}, at least one pair of nodes will not match nodes with the same initial labels, leading to a distance greater than $0$ for nodes $x \not\equiv y$.

\subsubsection{Proof of Symmetry}
\label{sec:symmetry}
The condition of symmetry requires that the proposed distance function must satisfy the property $d(x,y)=d(y,x)$. We can prove that Equation \ref{eq:decomposition} is symmetric, as

\begin{align*}
\begin{split}
d(\ell_{\ilast}(x),\ell_{\ilast}(y))&=
\sum_{(x',y') \in M'(x,y)} \vert deg(x')-deg(y') \vert \\
&=\sum_{(y',x') \in M'(y,x)} \vert deg(y')-deg(x') \vert\\ &= d(\ell_{\ilast}(y),\ell_{\ilast}(x))
\end{split}
\end{align*}

\noindent since $M'(x,y)$ is equal to $M'(y,x)$ when we swap the nodes. The order of the nodes in each pair does not affect the result of our distance function.

\subsubsection{Proof of the Triangle Inequality}
The triangle inequality requires that the inequality $d(x,y) \leq d(x,z)+d(z,y)$ is satisfied by the proposed automorphic distance function. 

By the definition of the proposed distance, we know that

\begin{align*}
\begin{split}
d(x,y) &\leq d(x,z)+d(z,y) \implies\\
d(\ell_{\ilast}(x),\ell_{\ilast}(y)) &\leq d(\ell_{\ilast}(x),\ell_{\ilast}(z))+d(\ell_{\ilast}(z),\ell_{\ilast}(y)) \implies\\
\sum_{(a,b) \in M'(x,y)} \vert deg(a)-deg(b) \vert &\leq
\sum_{(a,c) \in M'(x,z)} \vert deg(a)-deg(c) \vert 
+ \sum_{(c,d) \in M'(z,y)} \vert deg(c)-deg(d) \vert.
\end{split}
\end{align*}

We know that the absolute value satisfies the triangle inequality, thus the lower value that the right side can take is

\begin{align*}
\begin{split}
\sum_{(a,b) \in M'(x,y)} \vert deg(a)-deg(b) \vert \leq \sum_{(a,d) \in M''(x,y,z)} \vert deg(a)-deg(d) \vert
\end{split}
\end{align*}

\noindent where $M''(x,y,z)$ is the set of pairs resulting from chaining or combining $M'(x,z)$ and $M'(z,y)$ so that $(a,d) \in M''(x,y,z)$ if and only if, $(a,c) \in M'(x,z)$ and $(c,d) \in M'(z,y)$.

For this inequality to hold, it requires the non-existence of a pairing of nodes better than the matching done in the left side. Since the Hungarian algorithm ensures that matchings are optimal, minimizing their sum of distances, the matching at the right side cannot be better than optimal and, therefore, the value of the right side can only be equal to or greater than the value on the left side, satisfying the triangle inequality condition.

\subsection{Metric Computational Complexity}
In this section, we analyze the temporal and spatial computational complexity of our proposed metric.

The initialization of labels based on the degree of nodes has $O(n)$ temporal and spatial complexity, where $n$ is the number of nodes in the network. The computation of the table for the initial distances has $O(n^2)$ temporal and spatial complexity, since the distance is computed for every pair of nodes. 

Each iteration of the algorithm requires computing the sorted list of labels of the neighbors for each node. This task can be accomplished for each node with computational and spatial complexity $O(k)$, where $k$ is the degree of the node, when using radix, bucket, or counting sort. Thus, computing these strings for all nodes has $O(nk)$ temporal and spatial complexity. Renaming these labels can be done in $O(n)$ time using hash-based data structures. To compute the pairwise distances between labels in the current iteration, the Hungarian algorithm, with computational complexity $O(k^3)$, must be computed for each pair of nodes, leading to $O(n^2k^3)$ temporal complexity and $O(n^2)$ spatial complexity. Finally, checking if the labels have stabilized can be done in $O(n)$ using a hash-based index.

The number of iterations, $m$, required for 1-dimensional Weisfeiler-Lehman algorithm to converge is closely related to the diameter of the network \citep{furer2001weisfeiler}. Even though the number of iterations is theoretically bounded by $n$, it has been widely observed that real-world networks tend to exhibit the small-world phenomenon, presenting a small diameter \citep{travers1967small, watts1998collective} and leading to a small number of iterations required for convergence. 

Therefore, by combining these partial results, the total temporal complexity is $O(n+n^2+mn^2k^3) \approx O(mn^2k^3)$, where $n$ is the number of nodes in the graph, $k$ is the degree of nodes, and $m$ is the number of iterations required for convergence. The spatial complexity of the algorithm is $O(n^2)$, since only the distances and labels from the previous and the current iteration must be maintained at any given time. In practice, the algorithm can handle large networks, given that $m$ and $k$ tend to remain small in real-world networks due to their sparse and small-world nature. In addition, most of these steps can be easily parallelized, since most of them are independent for each node and are only based on the results from the previous iteration. For example, the initial labels for each node can be assigned independently. Once we have assigned these labels, the computation of their pairwise distances can be split among the available processors, since they are independent. The iterative assignment of labels in each loop iteration can be also parallelized using a concurrent data structure to ensure that the new labels are properly generated. Furthermore, the pairwise distances between these new labels can be easily computed in a parallel way, since they are completely independent as they only rely on the distance table computed in the previous iteration.

\section{Experimental Evaluation}
The experimental evaluation of role discovery techniques is a complicated task due to the lack of available evaluation datasets. Most role mining research projects use private datasets, which are not released and made publicly available to the scientific community. To overcome this limitation and perform an illustrative comparison of automorphic distances, we propose a novel experimental evaluation, which is presented in this section. As a collateral result of this experimentation, a new method for representing node roles using feature vectors is presented.

A central problem in machine learning is finding representations that ease the visualization or extraction of useful information from data \citep{bengio2013representation}. A common solution is the computation of embeddings that represent complex objects in a vector space preserving certain properties \citep{goldberg2014word2vec}. Node embedding, also known as graph embedding, is the task of mapping each node in a graph to a dimensional space trying to preserve the similarity or distance between pairs of nodes. Therefore, similar nodes will be located in similar regions of the space. Node embeddings have lately gained attention since they have achieved good results in different machine learning tasks \citep{goldberg2014word2vec}. Several models have been proposed for node embedding. However, these techniques try to preserve the connectivity of the network by obtaining embeddings that preserve structural equivalence, the neighborhood, or the community of nodes \citep{perozzi2014deepwalk,grover2016node2vec,zheng2016node}. This information has proven to be useful due to the presence of homophily, also known as assortativity, in real-world networks \citep{mcpherson2001birds}, where entities tend to be connected to similar ones, a feature that allows us to explain certain features of the nodes. Even though the connectivity information captured by these techniques is relevant, these techniques fail to capture information related to the role or function of the nodes in the network, which is a highly-valuable information that is complementary to the information obtained by locality-based embedding techniques.

We propose a new kind of embedding by exploiting this complementary information. Our node embeddings capture the roles of nodes by placing nodes that play a similar function in the network close in the resulting vector space. To compute these embeddings, we apply the classical multidimensional scaling (MDS) \citep{wickelmaier2003introduction,borg2005modern} to the distance matrices computed by the different approaches. In the following section, we show how 2-dimensional embeddings can capture relevant information in different real-world networks. We compare the results obtained by our distance metric with the results obtained by RoleSim, which can be interpreted as a distance function as previously described.

\subsection{Zachary's Karate Club Network}

Zachary's karate club network is a popular social network representing the $34$ members (as nodes) of a university karate club and their interactions (as edges) outside the club \citep{zachary1977information}. During the study carried out by Wayne W. Zachary, a conflict arose between the two club administrators, leading to the split of the club into two groups according to the leader each member decided to follow. For this reason, this network has served as a prototypical case study for community detection algorithms and some network analysis techniques.

If each member is assigned to a binary class according to the leader it decided to follow, Zachary observed that this property is highly homogeneous and assortative. Therefore, nodes tend to be connected to nodes that took the same decision. In this context, previously proposed embedding techniques generate embeddings that clearly separate nodes according to the leader they decided to follow \citep{perozzi2014deepwalk}. This information is crucial when the task is related to community detection. However, these embeddings fail, by nature, to reveal the role of each node in the network.

\begin{figure*}[!ht]
\centering
\begin{subfigure}[t]{0.29\textwidth}
\includegraphics[width=1\textwidth]{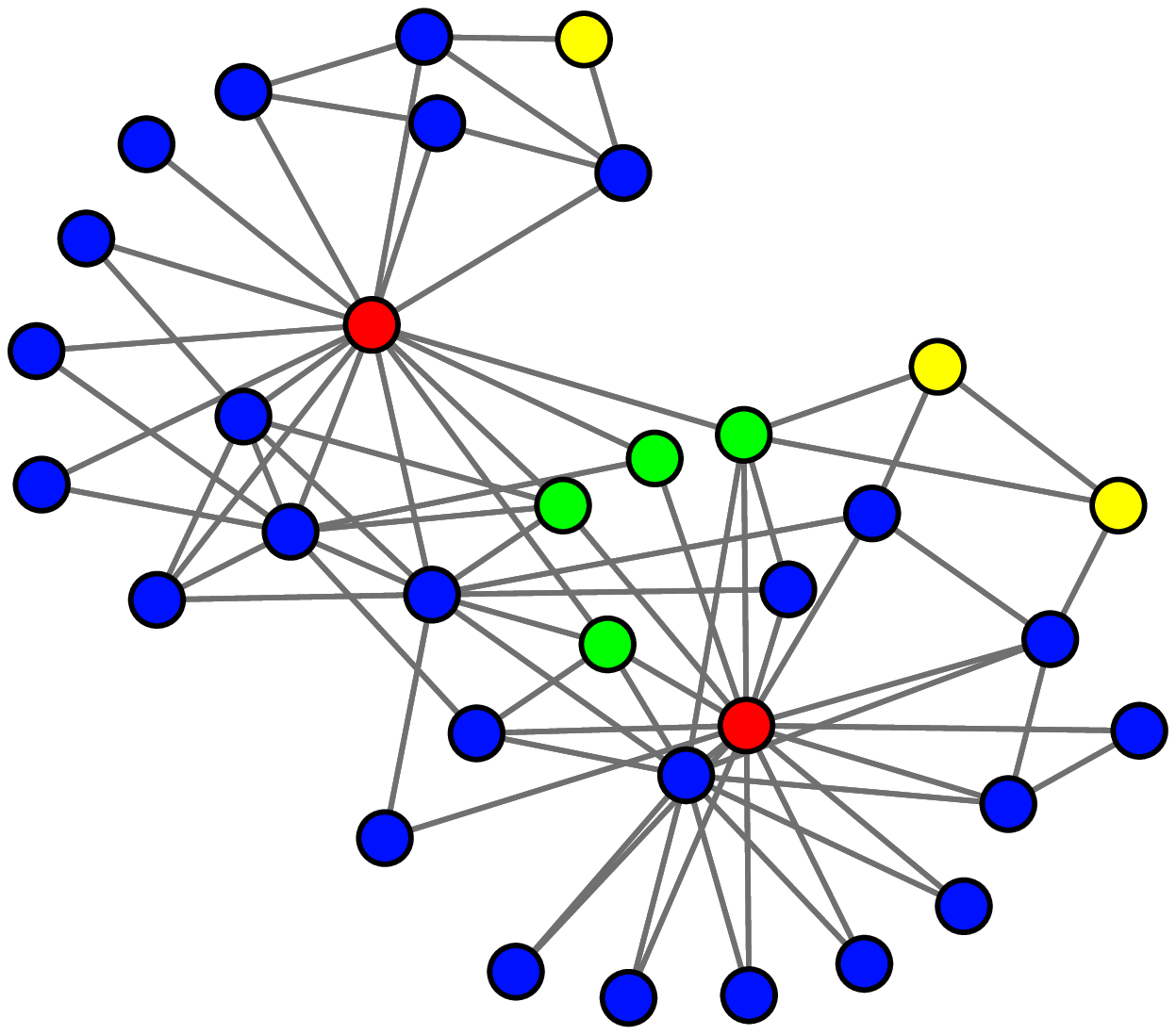} 
\caption{Kamada-Kawai layout.}
\end{subfigure}
\begin{subfigure}[t]{0.29\textwidth}
\includegraphics[width=1\textwidth]{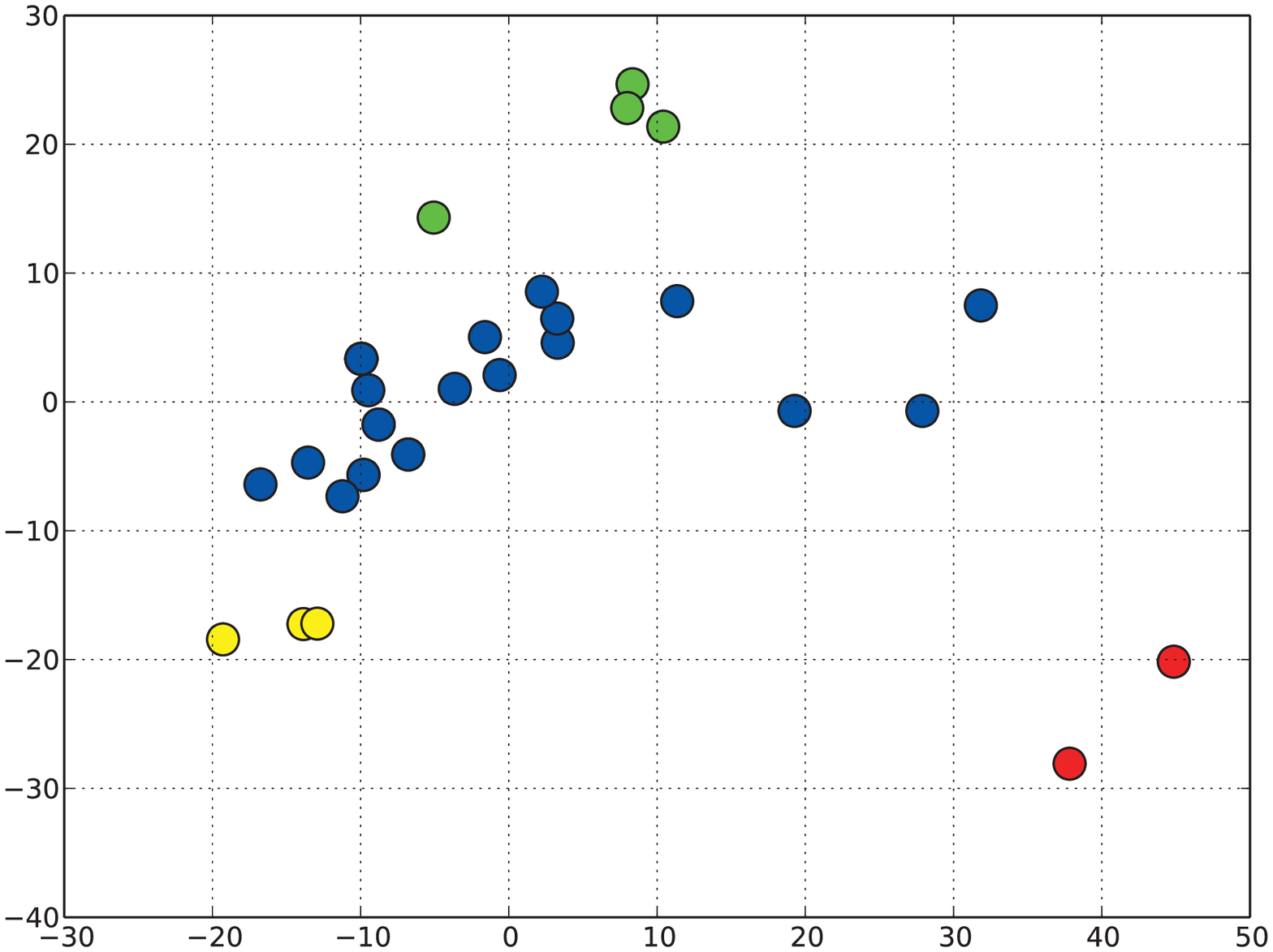} 
\caption{Automorphic embedding.}
\end{subfigure}
\begin{subfigure}[t]{0.29\textwidth}
\includegraphics[width=1\textwidth]{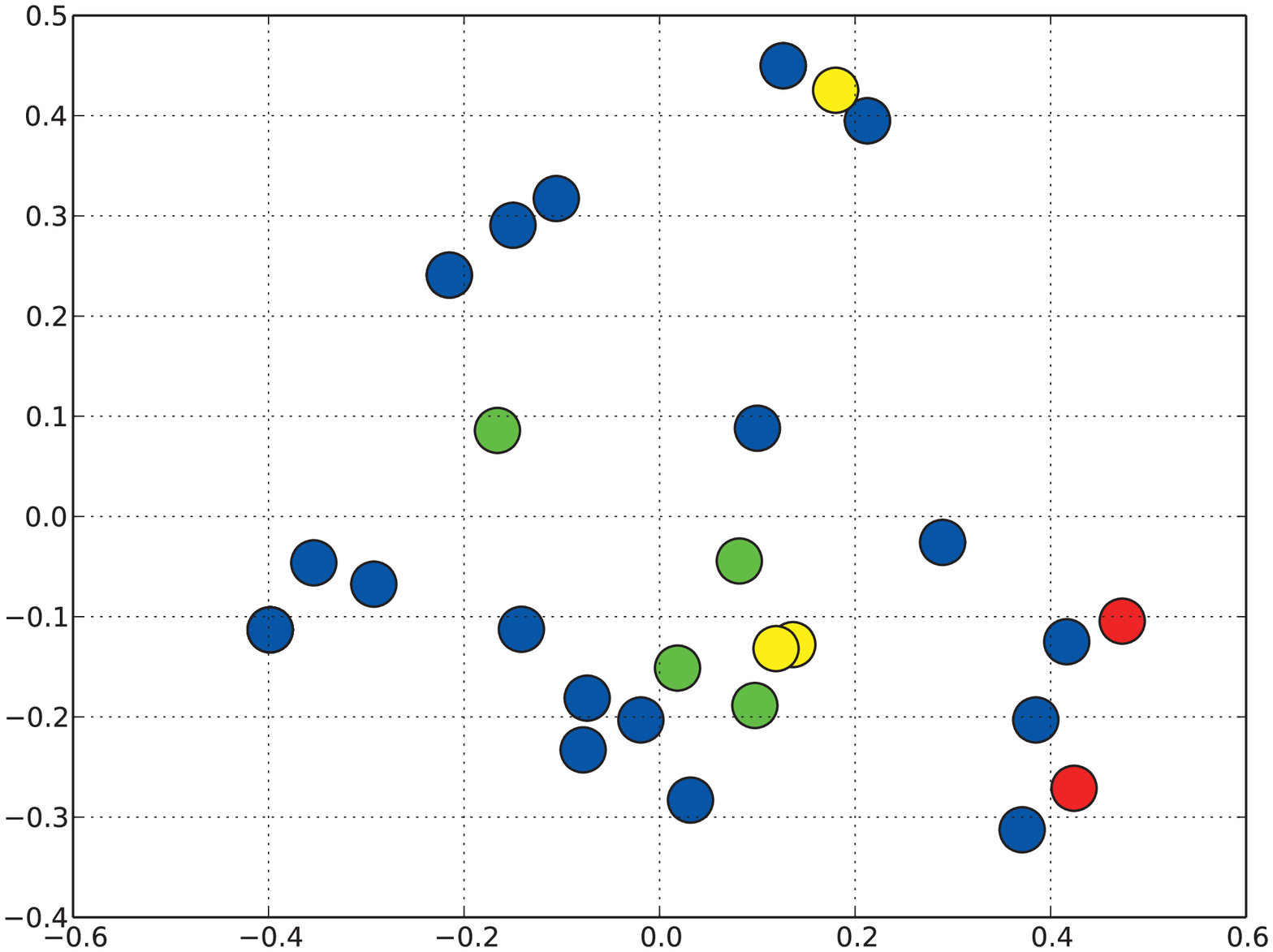} 
\caption{RoleSim embedding.}
\end{subfigure}
\caption{Zachary's karate club network and its node embeddings, with node role coloring.}
\label{fig:karate}
\end{figure*}

To illustrate how our technique captures the roles of nodes, we assigned a class to each node according to objective role-related properties. The two leaders are colored in red. Nodes interacting with the two leaders are colored in green. Nodes not interacting with any of the two leaders are colored in yellow. Finally, the remaining nodes, which interact with only one of the leaders, are colored in blue. Figure \ref{fig:karate} shows the network drawn using the Kamada-Kaway layout algorithm \citep{kamada1989algorithm} and the embeddings obtained by applying our automorphic distance and RoleSim. It can be seen the four classes of roles are linearly separated in the embedding generated using our distance metric. In addition, it can be seen how the two leaders are clearly mapped as outliers, placed significantly apart from the other nodes representing normal members of the club. However, RoleSim fails to clearly separate the different role classes. The two leaders are placed pretty close to normal club members, without capturing their unique function in the network.

\subsection{World Trade Network}
In network data mining, homophily is commonly exploited in node classification tasks, since nodes tend to exhibit the same class than their neighbors \citep{bhagat2011node}. Even though this situation occurs in a large number of networks from very different domains, homophily-based classification techniques fail when the classes of the nodes are defined by the role they play in the network, instead of by the community they belong to.

An illustrative example of this situation is a network containing data on trade of miscellaneous manufactures of metal among 80 countries \footnote{The dataset can be downloaded from \url{http://vlado.fmf.uni-lj.si/pub/networks/data/esna/metalWT.htm}}, according to data gathered in 1993 and 1994 from the Commodity Trade Statistics published by the United Nations \citep{de2011exploratory}. Each country is represented by a node in the network. Each commercial relationship is represented by an arc, which we consider an undirected edge in practice. In this case, arcs correspond to trading high technology products or heavy manufactures between countries.

In addition, the authors of this dataset annotated countries in the network with their structural world economic position in 1994. World economic positions are a classification of countries in the context of the world-system theory that explains some complex dynamics observed in the real world \citep{chirot1982world, smith1992structure}. This classification splits countries into three possible categories: core countries (colored in green), semi-periphery countries (colored in blue), and periphery countries (colored in red). In short, core countries have a high economical, military, and political power, which allows them to control the world economic system. The periphery is composed of less developed countries, owning a disproportionately small share of global wealth. Finally, the semi-periphery is conformed by countries that do not clearly fall in the previous two categories and exhibit a more intermediate status.

\begin{figure*}[!ht]
\centering
\begin{subfigure}[t]{0.29\textwidth}
\includegraphics[width=1\textwidth]{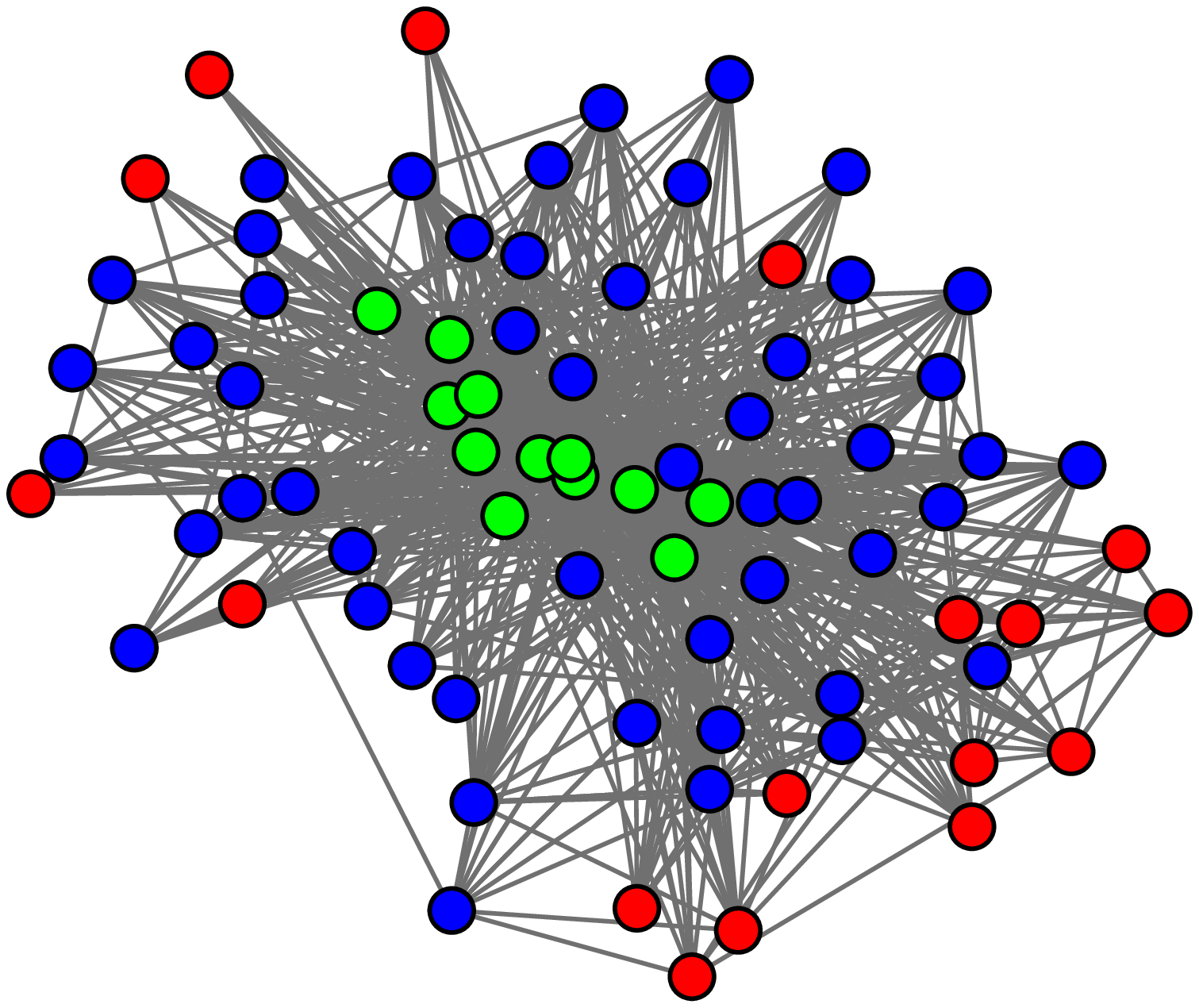} 
\caption{Kamada-Kawai layout.}
\end{subfigure}
\begin{subfigure}[t]{0.29\textwidth}
\includegraphics[width=1\textwidth]{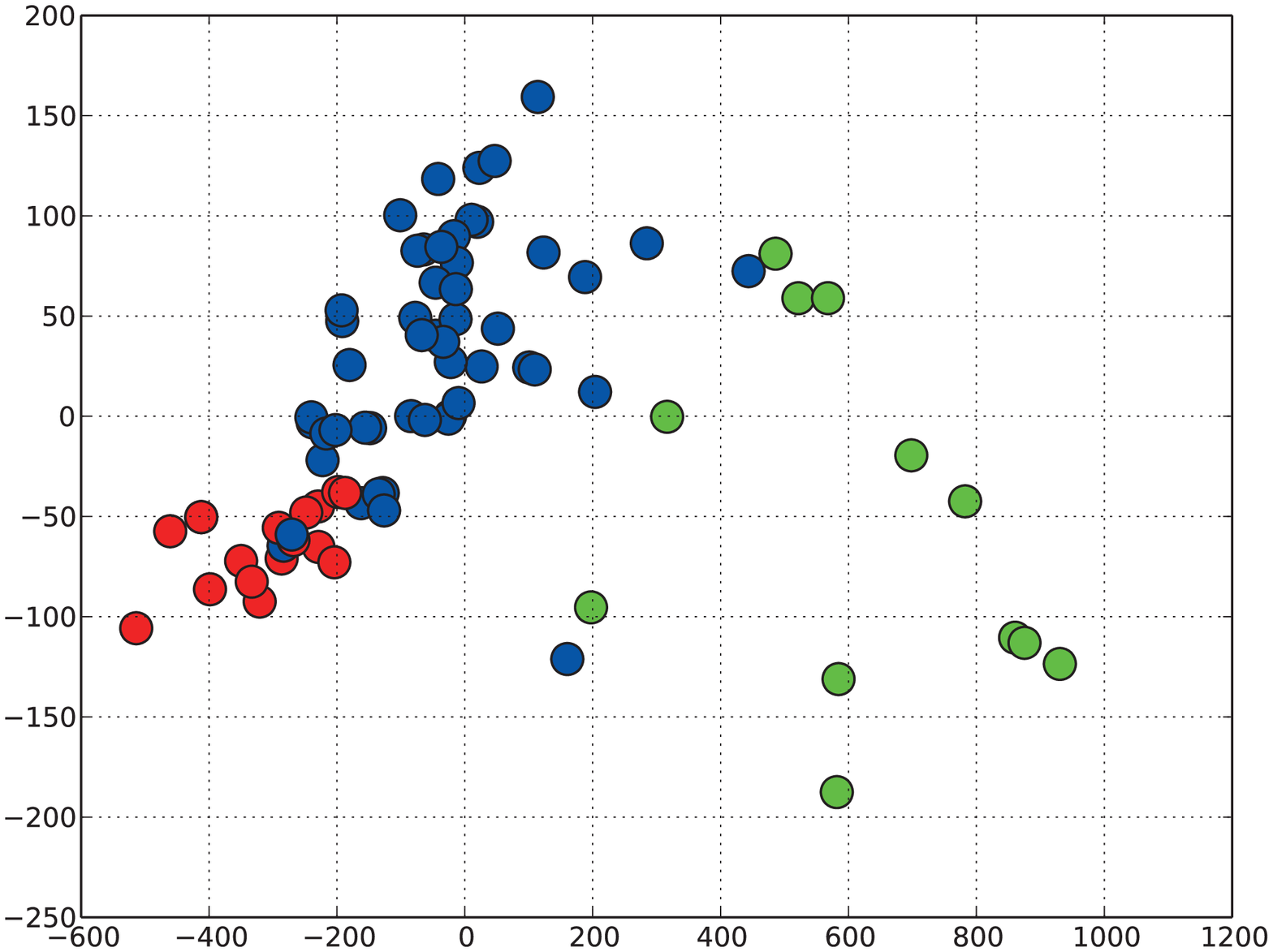} 
\caption{Automorphic embedding.}
\end{subfigure}
\begin{subfigure}[t]{0.29\textwidth}
\includegraphics[width=1\textwidth]{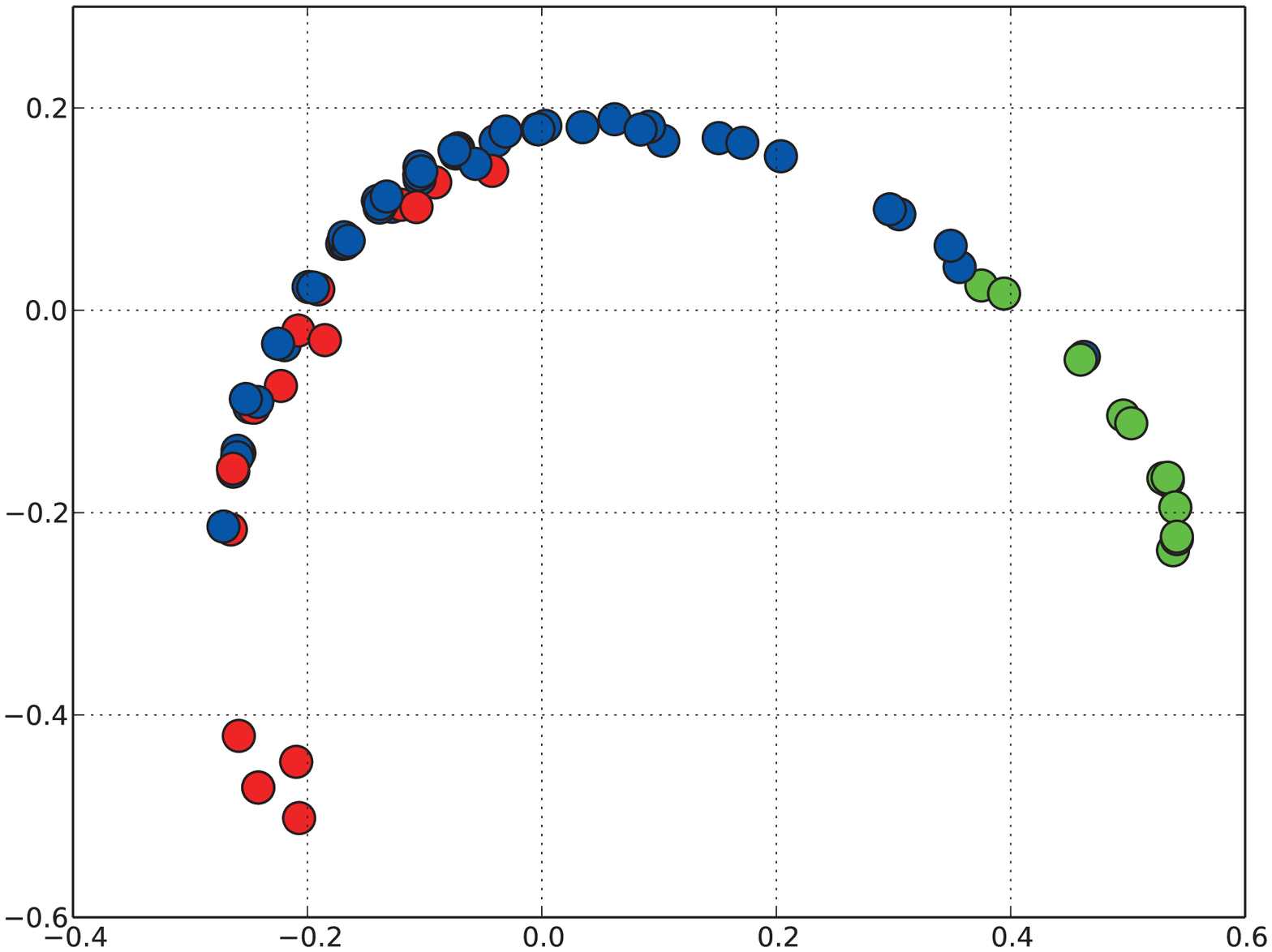} 
\caption{RoleSim embedding.}
\end{subfigure}
\caption{World trade network and its corresponding node embeddings.}
\label{fig:worldtrade}
\end{figure*}

Figure \ref{fig:worldtrade} shows the trade network drawn using the Kamada-Kaway layout algorithm and the embeddings obtained by applying our automorphic distance and the RoleSim-based distance. It can be seen that both metrics achieve a good separation of core and semi-periphery countries. However, our distance metric is clearly superior in the separation of semi-periphery and periphery countries. In addition, the embedding generated using distances computed with RoleSim exhibits an artificial curved-line shape, which indicates that only a dimension would be required to represent the information captured by RoleSim. In contrast, the embedding generated using our proposed automorphic distance exploits the two dimensions to represent role-related node properties and achieves a better separation of the different classes.

\section{Conclusions}
In this paper, we have proposed a novel distance metric for nodes that relaxes the strict concept of automorphic equivalence. To the best of our knowledge, this is the first work to propose a consistent non-normalized distance metric that captures the concept of automorphic equivalence without approximating it using feature engineering. In addition, we have shown that the proposed distance function is a valid distance metric by proving the required conditions.  Finally, we have shown how our metric can be exploited to generate node embeddings that capture role information in contrast to previously proposed embedding techniques, which capture locality-based information. We have also shown how our metric is superior to RoleSim in the generation of automorphic node embeddings, leading to a better separation of nodes according to their roles.

Our proposal creates new opportunities in problems related to role discovery. Future work includes exploiting our distance metric in problems related to anomaly detection in networks and transfer learning based on roles shared by nodes across different networks.

\section*{Acknowledgments}
This work is partially supported by the Spanish Ministry of Economy and the European Regional Development Fund (FEDER), under grant TIN2012-36951 and the program ``Ayudas para contratos predoctorales para la formaci\'on de doctores 2013'' (grant  BES-2013-064699).

\bibliographystyle{natbib}
\bibliography{article}

\end{document}